# Interlayer charge transfer from contact electrification in conducting micro and nanoscale thin film heterostructures

Sandeep Kumar[1,*] and Ravindra G Bhardwaj[2]

[1]Independent researcher, Kurukshetra, India

[2] Department of Mechanical Engineering, Birla Institute of Technology and Science Pilani, Dubai Campus, Dubai, the United Arab Emirates

[*] Corresponding author

Email: sandeep.suk191@gmail.com

**Abstract**

Contact electrification give rise to charge accumulation at the interface when two materials are brought into contact with each other. The charge accumulation at the interface will diffuse to the interior of the conducting material if the dimensions of the contacting conducting material is of the order of an unknown critical length scale. This contact electrification induced interlayer charge transfer will modify the fundamental physical properties of both the contacting materials. This review first discusses the reported experimental evidence of flexoelectricity induced contact electrification and interlayer charge transfer in conducting thin film based heterostructures. The interlayer charge transfer creates a gradient of charge carrier in both the thin films constituting the heterostructure and also modifies the electron-electron interactions. Further, the interlayer charge transfer changes the electron-phonon coupling, spin-phonon coupling and magnetoelectronic coupling that give rise to new physical behavior, which did not exist prior to the interlayer charge transfer. The new physical behaviors from interlayer charge transfer and their mechanistic origins are reanalyzed and discussed, which include spin-Hall effect of charge carriers, topological Hall effect of magnetoelectronic electromagnon, inhomogeneous magnetoelectronic multiferroic effect, flexoelectronic proximity effect and topological spin texture. This review article presents a unified picture of current status and future directions that will provide the scientists a stepping stone for research in the field of flexoelectricity mediated contact electrification and interlayer charge transfer mediated behavior in the micro/nanoscale heterostructures of the conducting materials.

1. **Introduction**

When two solids are brought into contact with each other, an interlayer charge transfer may occur due to the difference in electronic properties. One of the well-known examples of this behavior is contact electrification (CE). In the CE, charge arises at the surfaces of two materials when they are brought into contact with each other and then separated[1-5]. The CE behavior is observed across all kind of solids (metals, insulators, semiconductors and polymers). The CE behavior is, possibly, a combined effect due to difference in electronic properties[1, 5], curvature of interacting surfaces[1, 6], flexoelectricity[7, 8], material transfer[9] and friction[1, 8] among others. The CE occurs even if the materials are brought into contact but not subsequently separated from each other[3]. This interfacial charge accumulation can give rise to two-dimensional electron gas (2DEG) especially in case of heterostructure of two dielectric materials[10-12]. Further, the properties of 2DEG can also be controlled using flexoelectricity[13-15], which is, again, a subset of the CE behavior. In the two dimensional limit, the difference in electronic properties in 2D material heterostructures[16] lead to ultrafast charge transfer[17, 18], which has been explored for photogeneration[19] and catalysis[16] among other applications.

The CE response has been extensively reported in conducting materials and charge accumulation occurs until their Fermi levels are coincident[3, 5, 20]. Considering bulk materials, the CE induced charge accumulation remains localized at the surface (or interacting interface) even in the case of conducting materials as shown in Figure 1(a). The surface/interface charge accumulation is screened by the large and delocalized charge carrier concentration in the bulk of the conducting material. As a consequence, the effect of the surface charge is not observed in the bulk properties. However, this may not be the case in conducting micro/nano scale thin film heterostructures especially if the CE is significant as shown in Figure 1 (b).

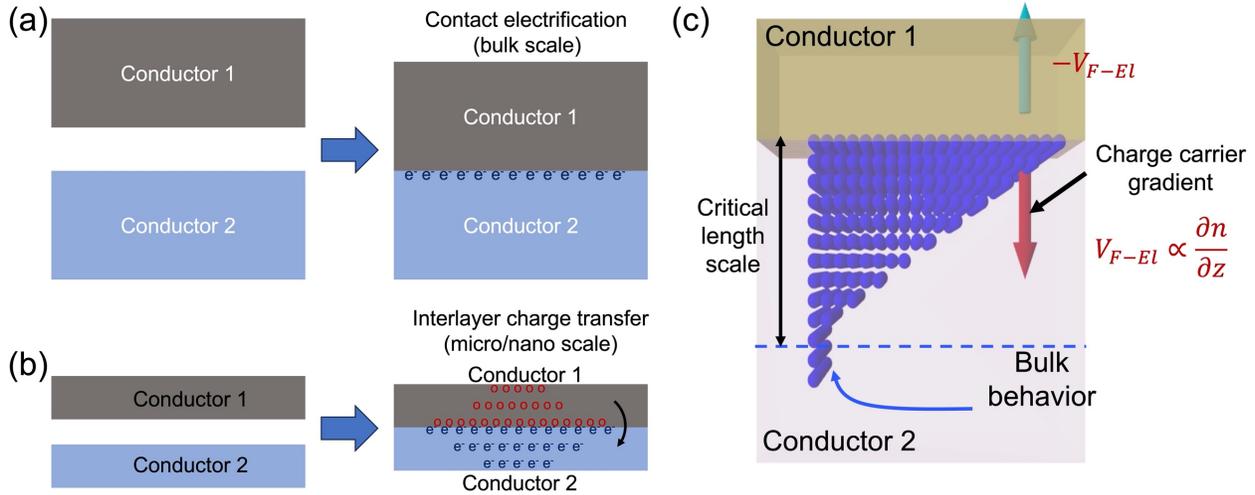

Figure 1. Schematic showing (a) the contact electrification in bulk conductors and (b) interlayer charge transfer from contact electrification at micro/nano scale. (c) Schematic showing diffusion of charge carrier to a critical length scale leading to a gradient of charge carrier from contact electrification induced interlayer charge transfer. ((c) Adapted and reproduced with permission from reference [21]. Copyright (2022) by the American Physical Society)

The CE induced charge accumulation in micro/nano scale conducting thin film systems will diffuse away from the interface/contact surface to a critical length scale as shown in Figure 1 (c), which is still unknown. If the dimension of the thin film is of the order of the critical length scale, then, charge diffusion away from the interface will give rise to a gradient of charge carrier as shown in Figure 1 (b,c). This transfer of charge across the thickness of the thin film is called CE induced interlayer charge transfer. New physical and material behavior will emerge from interlayer charge transfer that did not exist prior to it. Hence, CE induced interlayer charge transfer in conducting systems can potentially provide an opportunity to uncover new materials physics and engineering. It is noted that triboelectrification behavior[22] related scientific studies are not considered here because only physical contact (using material growth and thin film deposition) without separation is relevant for interlayer charge transfer in this review. Currently, the scientific studies in the CE induced interlayer charge transfer in conducting micro/nano scale systems are sparse and unorganized. This review article addresses this deficiency by summarizing and organizing the current status of research pertaining to CE induced interlayer charge transfer in conducting systems. This review article first explores the reported experimental evidence[21, 23-25] of flexoelectricity induced CE and interlayer charge transfer. The interlayer charge transfer gave rise to electronic dynamical multiferroicity[23], interlayer coupling[26] and proximity effects[26]. These leads to altered physical properties such as spin-phonon coupling[23, 27], electron-phonon coupling and magnetoelectronic coupling[25], which are underlying cause of phonon magnetic moment[23], spin-Hall effect from phonon skew scattering[23, 28], magnetoelectronic electromagnon excitations[24], inhomogeneous magnetoelectronic multiferroic effect[24] and topological spin texture[26]. The second part summarizes these new physical behaviors and also presents a structured reanalysis of the reported materials behavior in the context of interlayer charge transfer. Lastly, this review list challenges and directions for future research.

## 2. Evidence of interlayer charge transfer in conducting heterostructures

The experimental measurement of CE induced interlayer charge transfer in conducting thin films heterostructures is a challenge. In the bulk metal samples, the charge accumulation is directly measured using electrometer[20], which is difficult to implement for thin film heterostructures. In dielectric thin films, the charge accumulation at the surface due to the CE can be studied using

Kelvin probe force microscopy (KPFM)[7, 29] but hasn't been applied to conducting thin film heterostructures. The transport measurement is another technique that has been used to study formation of 2DEG in dielectric material systems. At micro/nano scale, the CE induced interlayer charge transfer was successfully studied using transport measurements in case of conducting thin film heterostructures. The CE induced interlayer charge transfer will increase the carrier concentration in one layer and reduce in the other, which will change the resistances of both the layer constituting the heterostructure. The conductivity of the heterostructure thin film sample is expected to be different from the value based on resistors in parallel configuration, which can be used to quantify the interlayer charge transfer. Similarly, the change in charge carrier concentration can also be studied using the Hall effect measurements, which can be used to understand interlayer charge transfer.

One of the primary mechanisms of CE is flexoelectricity as stated previously[7, 8]. As compared to other mechanisms (curvature of interacting surface, material transfer and friction etc.) strain gradient required for flexoelectric effect can be achieved easily using two methods, one, a freestanding thin film structure at micro/nanoscale where residual stresses induce buckling[21, 28] of the thin film and second when the thin film sample is constrained/encapsulated[26, 30, 31] (along with mismatch in coefficient of thermal expansion of thin films). The freestanding structure was achieved using conventional micro/nano fabrication techniques. The fabrication method details are not discussed here and can be found elsewhere[28, 32, 33]. The experimental setup to uncover CE induced interlayer charge transfer had freestanding thin film in four probe and Hall bar configurations. Further, two different sample configurations have been reported in literature so far that shows the evidence of interlayer charge transfer in conducting heterostructures. In the first configuration, a metal/degenerately doped Si heterostructure[21] was reported. The degenerately doped Si behaves similar to a metal but with significantly smaller charge carrier concentration (~$10^{19}$ cm$^{-3}$) and larger resistivity (~$10^{-4}$-$10^{-5}$ Ωm). A combination of metal and degenerately doped Si had significantly different electronic properties, which gave rise to measurable interlayer charge transfer. From the KPFM experiments, we know that flexoelectricity give rise to CE between Pt tip and dielectric surface[7]. The second sample configuration had a dielectric (MgO)/degenerately doped Si heterostructure[23, 25]. The flexoelectricity mediated CE and interlayer charge transfer was reported in both the configurations, which are discussed below.

2.1 Metal-degenerately doped Si based heterostructures

Lou et al.[21] reported first experimental evidence of interlayer charge transfer in freestanding Permalloy (Py) (25 nm)/MgO (1.8 nm)/p-doped Si (400 nm) heterostructure sample [21]. The resistance measurement as a function of temperature from 350 K to 5 K at 10 µA of alternating current showed a Mott metal insulator transition (MIT) behavior below ~50 K as shown in Figure 2 (a). A residual resistance was observed below the Mott MIT as shown in Figure 2 (b), which showed that entire sample did not go through the MIT. The resistance response across the temperature range was found to be a function of applied current through the sample as shown in Figure 2 (a). At 350 K, the resistance of the sample was lower at larger current as shown in Figure 2 (a) whereas residual resistance was higher at larger current as shown in Figure 2 (b). This observed behavior was attributed to interlayer charge transfer where charge carrier transfer from Py to p-Si layer induced Mott MIT in the p-Si layer as shown in Figure 2 (c) whereas Py layer remaining an electrical conductor. Based on the resistivity of a Py only sample, the residual resistance of the sample should have been ~182.6 Ω at 5 K if there was no interlayer charge transfer. Instead, a resistance of ~ 245.8 Ω at 10 µA and 5 K was measured. The increase in the resistance of the Py was due to the interlayer charge transfer and resulting reduction in charge carrier concentration. The charge carrier concentration in Py and p-Si thin films were measured to be $6.85\times10^{22}$ cm$^{-3}$ and $8.73\times10^{18}$ cm$^{-3}$, respectively. But, the charge carrier concentrations in the heterostructure sample were estimated to be $5.08\times10^{22}$ cm$^{-3}$ and $1.11\times10^{21}$ cm$^{-3}$ for Py and p-Si layers at 10 µA current, respectively, based on the residual resistance as shown in Figure 2 (d). Similarly, the average charge carrier concentrations in the heterostructure sample were estimated to be $4.55\times10^{22}$ cm$^{-3}$ and $1.44\times10^{21}$ cm$^{-3}$ for Py and p-Si layers at 2 mA current, respectively, as shown in Figure 2 (d). Hence, the charge carrier concentration of Py reduced by $1.80\times10^{22}$ cm$^{-3}$ (26.3%) and $2.30\times10^{22}$ cm$^{-3}$ (33.6%) at 10 µA and 2 mA, respectively, as compared to Py only thin film. Similarly, the charge carrier concentration of p-Si layer increased by approximately three orders of magnitude due to interlayer charge transfer.

    An increase in the applied current led to Joule heating of the freestanding sample, which changed the thermal stresses as well as strain gradient (buckling of thin film structure) in the sample as shown in Figure 2 (d). As a consequence, the change in charge carrier concentration and interlayer charge transfer was attributed to the strain gradient (flexoelectricity) mediated CE as shown in Figure 2 (d). The reduction in resistance of the sample was due to increase in mobility of charge carrier in p-Si as compared to Py as shown in Figure 2 (d). At 300 K, the mobilities at 10 µA and 2 mA were estimated to be 3.88 cm$^2$/(V.s) and 10.61 cm$^2$/(V.s) [21], respectively. The original work called it flexoelectronic doping or flexoelectronic charge transfer, as shown in Figure 2(d), because it was interpreted as electronic response induced by the strain gradient present in the sample. However, this was an example of interlayer charge transfer from the flexoelectricity induced CE.

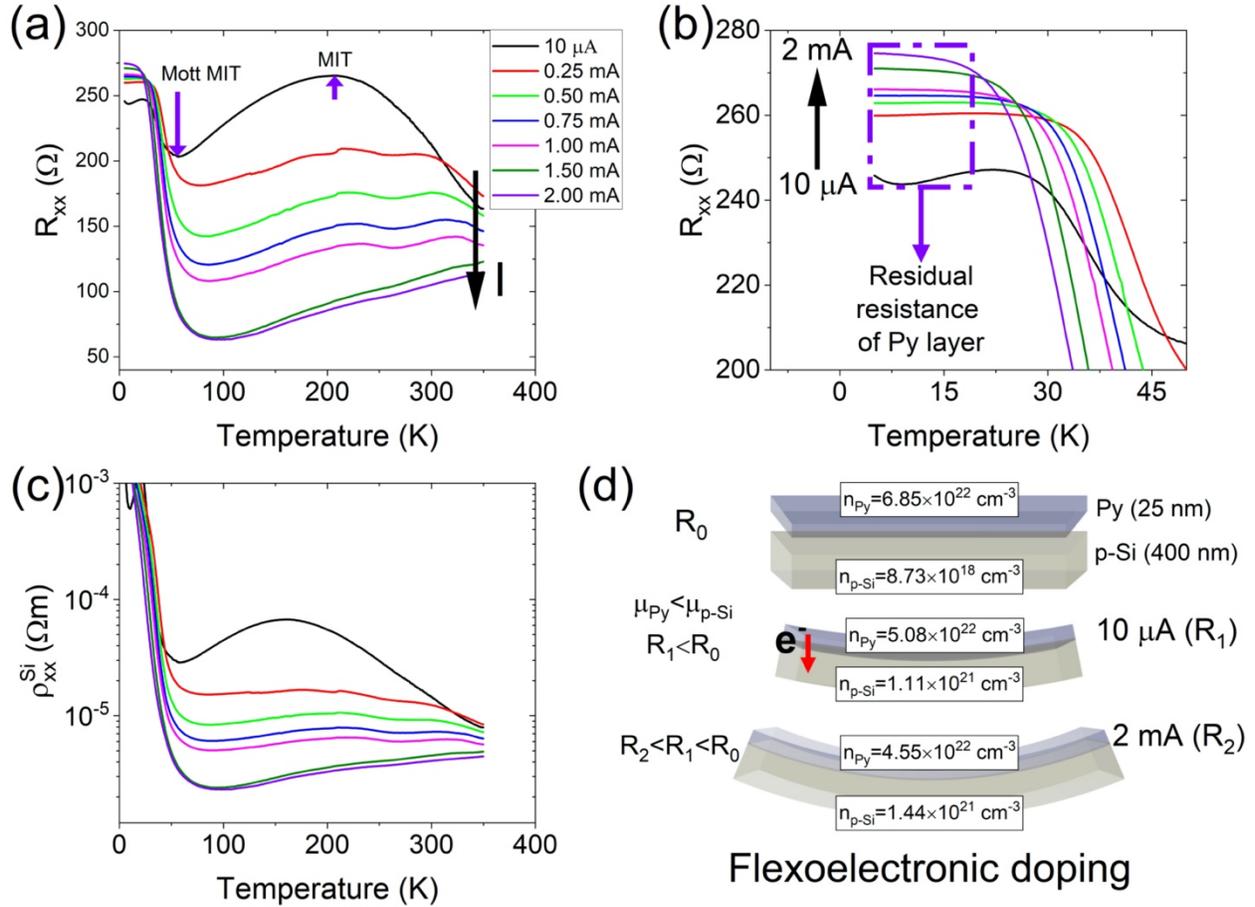

Figure 2. (a) The longitudinal resistance response in Py (25 nm)/MgO (1.8 nm)/p-Si (400 nm) composite sample as a function of temperature and current bias from 10 μA to 2 mA, (b) the resistance response between 50 K and 5 K showing the Mott metal insulator transition in the Si layer and residual resistance of Py layer, (c) the estimated resistivity of the Si layer in the composite sample after subtracting Py response, and (d) a schematic showing the change in the average charge carrier concentration from interlayer charge transfer from Py layer to Si layer. The $R_0$, $R_1$ and $R_2$ are sample resistances at no interlayer charge transfer, with interlayer charge transfer at 10 μA and 2 mA, respectively. (Reproduced with permission from reference [21]. Copyright (2022) by the American Physical Society)

As stated earlier, the Hall resistance measurement can also be used to study the interlayer charge transfer behavior[21]. Another study measured the Hall resistance in two samples: Py (25

nm)/MgO (1.8 nm)/p-Si (2 μm) and Py (25 nm)/MgO (1.8 nm)/p-Si (400 nm), as shown in Figure 3. The Hall resistance in the first sample having 2 μm p-Si was negative and a positive anomalous Hall resistance. Whereas, the second sample having 400 nm p-Si showed a positive Hall resistance and negative anomalous Hall resistance as shown in Figure 3. This measurement showed a large interlayer charge carrier transfer in the thinner sample due to larger buckling and strain gradient, which changes the sign of Hall and anomalous Hall resistances. This measurement reinforced the results from previous experiment. However, the quantitative information regarding charge carrier concentration and interlayer charge transfer was not estimated.

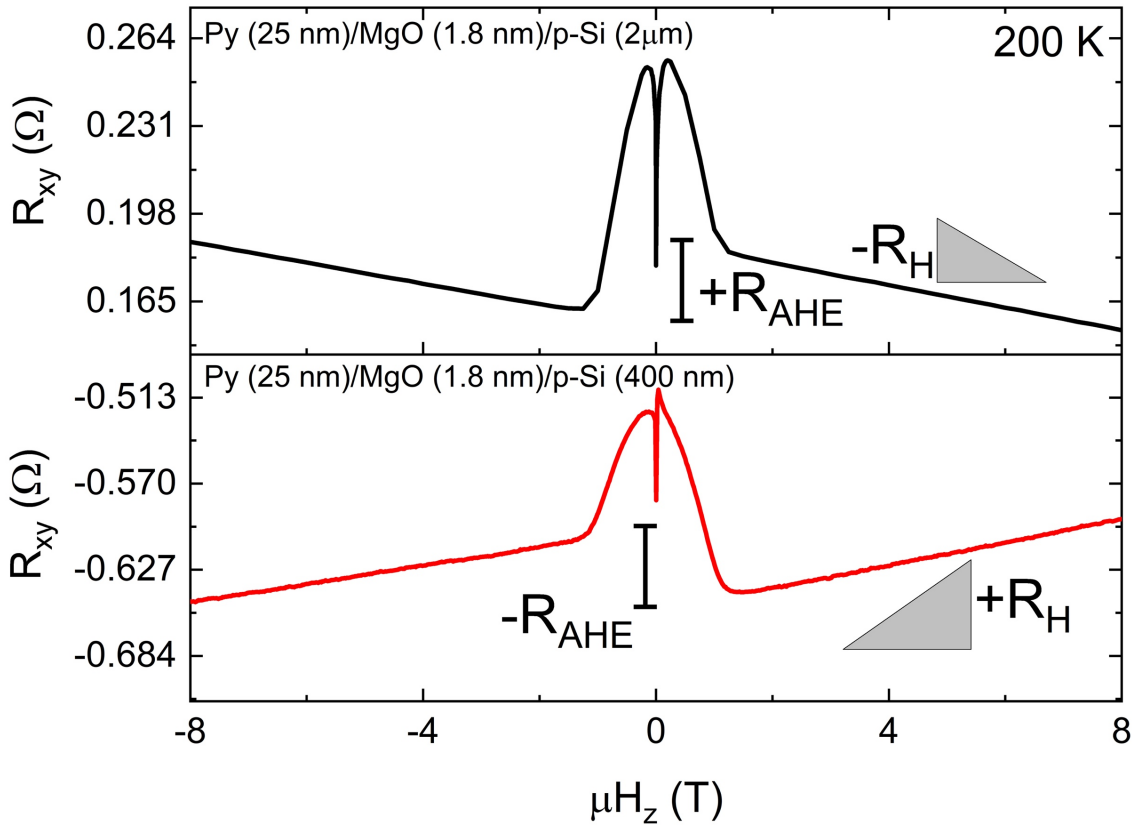

Figure 3. The Hall resistance measurement for an applied magnetic field of 8 T to -8T at 200 K showing negative Hall resistance in Py (25 nm)/MgO (1.8 nm)/p-Si (2 μm) in top panel and positive Hall resistance in Py (25 nm)/MgO (1.8 nm) /p-Si (400 nm) in bottom panel. (Adapted

and reproduced with permission from reference [21]. Copyright (2022) by the American Physical Society).

In the third experiment, the Pt (10 nm) was used instead of Py layer. In the Pt (10 nm)/MgO (1.8 nm)/p-Si (2 µm) sample, the longitudinal resistance and Hall resistance as a function of temperature were measured and reported[23]. In this sample, longitudinal resistance showed a transition between 300 K and 200 K as shown in Figure 4 (a). The, corresponding, Hall resistance measurement showed a change in sign attributed to change in carrier type as shown in Figure 4 (b). This behavior was again attributed to interlayer charge transfer due to change in strain gradient and CE.

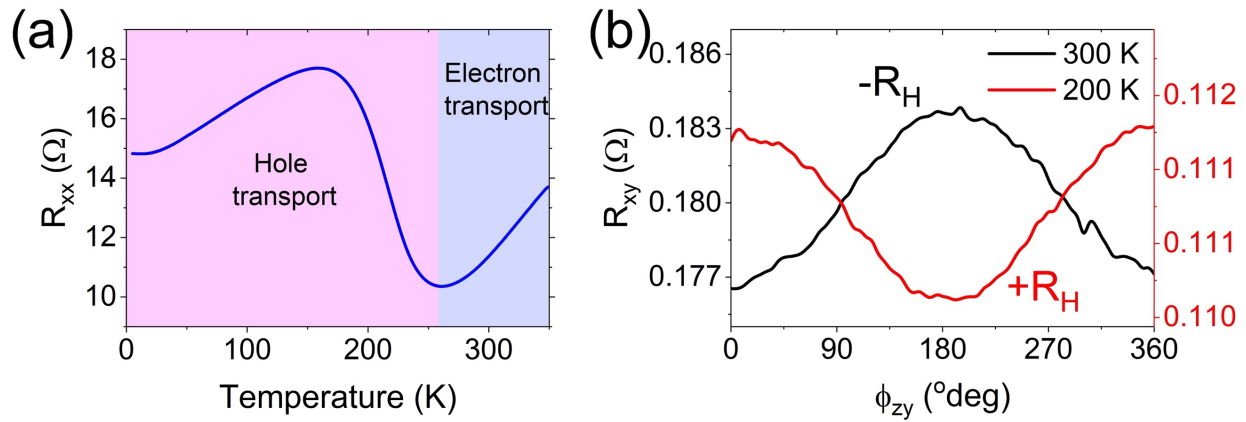

Figure 4. (a) the longitudinal resistance measurement as a function of temperature in Pt (10 nm)/MgO (1.8 nm)/p-Si (2 µm) sample from 350 K to 5 K. And (b) angle-dependent transverse resistance as a function of magnetic field (1 T) and temperature (300 K and 200 K) showing the sign reversal of Hall resistance. (Adapted and reproduced with permission from reference [23]. Copyright (2021) by the American Chemical Society).

The freestanding metal/degenerately doped Si thin film heterostructures buckle due to residual stresses from deposition and fabrication processing, which induces a through thickness strain gradient in the sample structure. The Figure 5 (a) shows a representative Si thin film structure that buckled due to residual stresses. Further deposition of the metal layer will induce even larger buckling. This was expected to cause an interfacial flexoelectric effect as shown in Figure 5 (b)[21, 24, 34]. Alternatively, this can also be considered an interfacial piezoelectric like effect[34]. The mismatch in the electronic properties and interfacial flexoelectric effect ($V_{FE}$) caused CE and interlayer transfer of charge carrier as shown in Figure 5 (c) that was observed in the three reported experimental studies. The interlayer charge transfer created a gradient of charge carrier with larger charge carrier concentration near the interface as shown in Figure 5(c). This gradient in charge carrier gave rise to flexoelectronic polarization ($P_{F-El}$) as shown in Figure 1(c) and 5 (c). In the first study, the strain gradient was modulated using Joule heating by passing larger current through the sample. In the second study, the thinner (400 nm) sample will have larger buckling and strain gradient than thicker (2 µm) sample. In the third study, the temperature change (thermal mismatch stresses) led to change in the strain gradient. These experiments presented evidence of interfacial flexoelectricity mediated CE[7, 8] and interlayer charge transfer in metal/ degenerately doped Si heterostructure thin film samples.

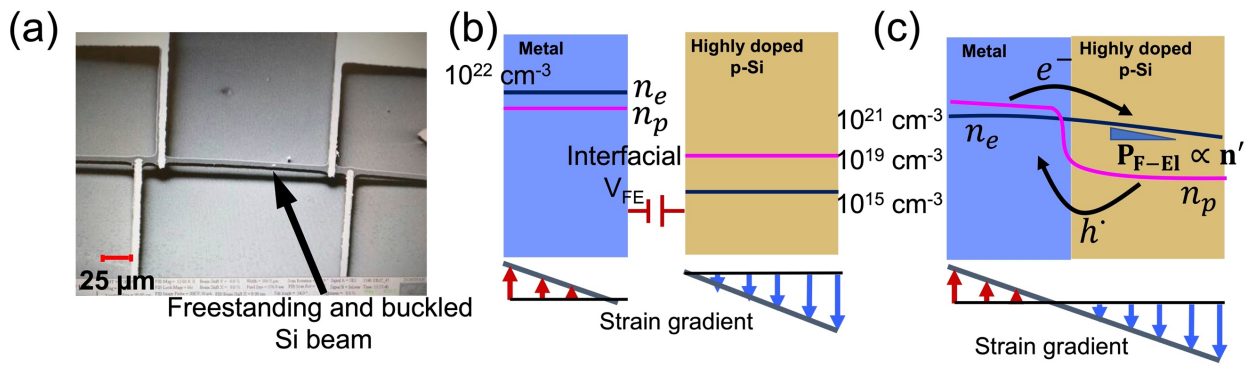

Figure 5. (a) A representative scanning electron micrograph showing the device structure for four-probe transport measurements having freestanding Si beam buckling due to residual stresses without any metal layer on top. (Adapted and reproduced with permission from reference [28].

Copyright (2020) by the American Physical Society). (b) An approximate distribution of charge carrier concentration in a metal/degenerately doped p-Si heterostructure and (c) a schematic showing the interlayer charge carrier transfer in doped Si from metal layer due to interfacial flexoelectric effect ($V_{FE}$) leading to gradient of charge carrier in Si and flexoelectronic polarization. (Adapted and reproduced with permission from reference [24]. Copyright (2023) by the American Physical Society).

2.1.1. Electron-electron interactions

The interlayer charge transfer leave one layer charge deficient and the other with excess charge. The first consequence of interlayer charge transfer was change in the electron-electron interactions. The change in electron-electron interactions (Coulomb repulsion) led to the Mott MIT response presented previously as shown in Figure 2. The Mott MIT also required magnetic exchange interactions in the p-Si layer, which were induced from the proximity effect. The evidence of the proximity effect will be discussed later in this article.

2.2 Dielectric-degenerately doped Si heterostructure

There have been two experimental reports that demonstrated CE behavior in dielectric/degenerately doped Si heterostructures. The first report used Hall effect measurement as a function of current in MgO (1.8 nm)/n-doped Si (2 µm) [23]. The Hall effect measurement at 1 mA of ac bias showed an estimated charge carrier concentration of ~$5.9 \times 10^{19}$ cm$^{-3}$ as shown in Figure 6(a). This value changed to ~$8 \times 10^{19}$ cm$^{-3}$ at 5 mA[23] of measurement current as shown in Figure 6(b). Similarly, the resistivity (resistance) of the sample reduced from ~$2.44 \times 10^{-5}$ Ω-m (34.95 Ω) at 1 mA to ~$1.5 \times 10^{-5}$ Ω-m (24.73 Ω) at 5 mA[23]. This study, originally, attributed the observed behavior to flexoelectronic charge separation.

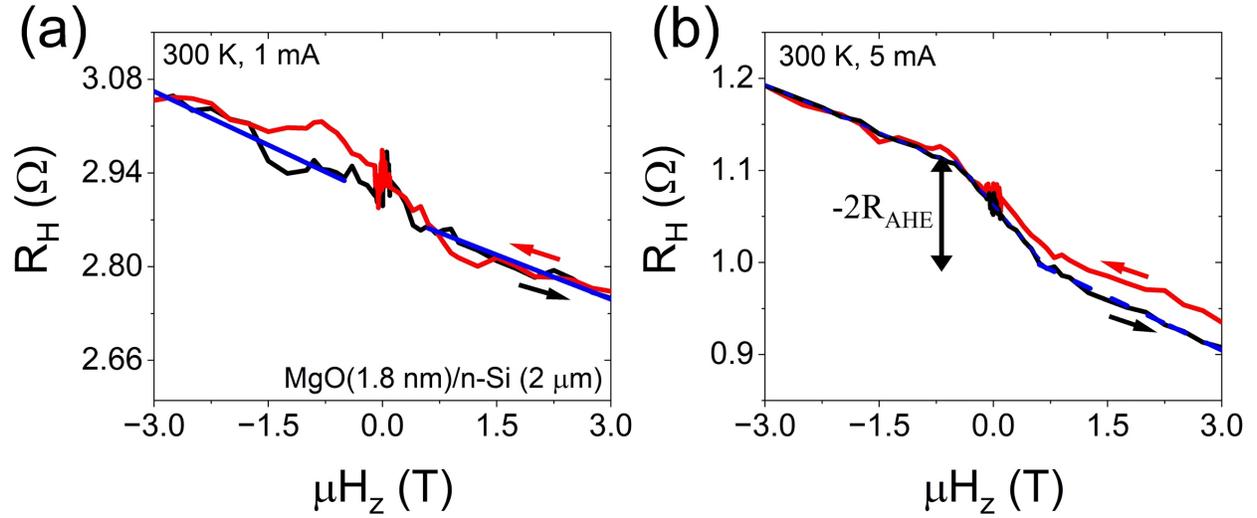

Figure 6. The anomalous Hall effect measurement in the MgO/degenerately doped n-Si (2 μm) sample at (a) 1 mA and (b) 5 mA of current. (Adapted and reproduced with permission from reference [23]. Copyright (2021) by the American Chemical Society).

In the second report, two sample with and without dielectric (control) were taken to uncover the interlayer charge transfer. Following were the configurations of the two samples: sample 1 having only p-Si (400 nm) (control sample) and sample 2 having MgO (2 nm)/p-Si (400 nm) heterostructure. In Hall effect measurement, the sample 1 showed a charge carrier concentration of $-3.84 \times 10^{18}$ cm$^{-3}$ whereas the charge carrier concentration was estimated to be $-9.7 \times 10^{18}$ cm$^{-3}$ for sample 2[25]. The difference in the charge carrier concentration between sample 1 and sample 2 was $4.86 \times 10^{18}$ cm$^{-3}$. The difference in the charge carrier concentration was, originally, attributed to the flexoelectronic charge separation.

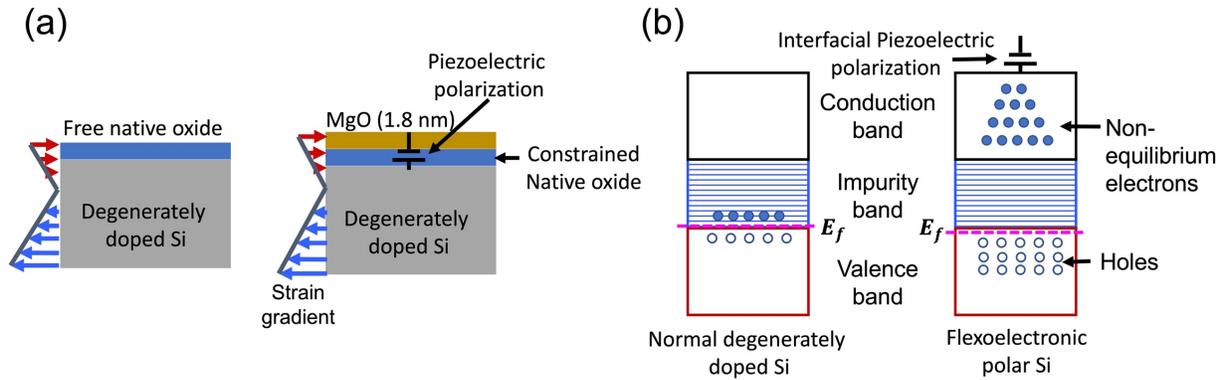

Figure 7. (a) Schematic showing the interfacial piezoelectric polarization in native oxide under an applied strain gradient, which acts on the degenerately doped Si layer and (b) schematic showing the charge carrier distribution due to interfacial piezoelectric like effect leading to non-equilibrium electrons in the conduction band giving rise to metastable flexoelectronic polar p-Si layer. (Adapted and reproduced with permission from reference [25]. Copyright (2024) by the American Physical Society).

The observed change in charge carrier concentration behavior in both dielectric/degenerately doped Si samples was similar to Metal/p-Si heterostructures except dielectric (MgO in the reported study) materials did not have the adequate charge carrier concentration to give rise to the observed behavior. As stated earlier the CE behavior arises due to a combination of mechanisms. In the case of dielectric/degenerately doped Si heterostructure, the CE behavior arose from strain gradient mediated piezoelectric like effect in the native oxide [34] on Si constrained by MgO layer as shown in Figure 7(a). The thickness of native oxide and MgO layers are significantly smaller than the thickness of Si layer and, as a consequence, strain across these layers is expected to be constant. Hence, the expected behavior is from piezoelectric like effect and not flexoelectric effect. However, the contributions from an interfacial flexoelectric effect as well as bulk flexoelectric effect cannot be ruled out. The resulting charge accumulation from piezoelectric effect at the interface gave rise to transfer of charge carrier from the impurity and valence band of Si to conduction band of Si as shown in Figure 7(b). Hence, this was an indirect interlayer charge transfer. These non-equilibrium charge carriers gave rise to change in Hall resistance and smaller resistivity measured in both the reported experimental studies. Similar to the metal/Si heterostructure, the non-equilibrium charge carrier concentration is expected to be larger at the interface as compared to farther away from the interface. This will give rise to a gradient in charge carrier concentration and flexoelectronic polarization. The charge neutrality is the primary difference between two material system. In the metal/Si heterostructures, the individual layers are no longer charge neutral due to interlayer charge transfer whereas charge neutrality is maintained in dielectric/Si heterostructures because non-equilibrium charge carrier do not violate it. The original studies did not explicitly mention the CE as the underlying cause in both material systems even though the mechanistic explanation was similar to flexoelectricity mediated CE[7, 8]. These studies have not only demonstrated the interlayer charge transfer behavior but it also showed that flexoelectricity is, most likely, the primary mechanism driving the electronic CE behavior in the bulk materials since removal of strain gradient led to absence of interlayer charge transfer[26].

2.3 Quality of interface

In conventional heterostructure materials, the quality of interface is essential to achieve required interlayer behavior. However, the CE induced interlayer charge transfer studies demonstrated that the quality of interface is not essential. The transmission electron micrograph of the Py/MgO/p-Si heterostructure showed that native oxide on the silicon was present in spite of etching attempts during fabrication as shown in Figure 8(a) [28]. This did not eliminate the interlayer charge carrier transfer even though the combined thickness of MgO and native oxide layer was ~5.5 nm as shown in Figure 8 (a) [28]. Hence, the strain gradient mediated interfacial flexoelectric effect can overcome traditional interfacial quality problems and facilitate CE induced interlayer charge transfer. In another Py/p-Si sample, the MgO and native oxide were absent as shown in Figure 8(b) and interlayer charge transfer mediated behavior was observed. However, the effect of MgO and native oxide was not quantified.

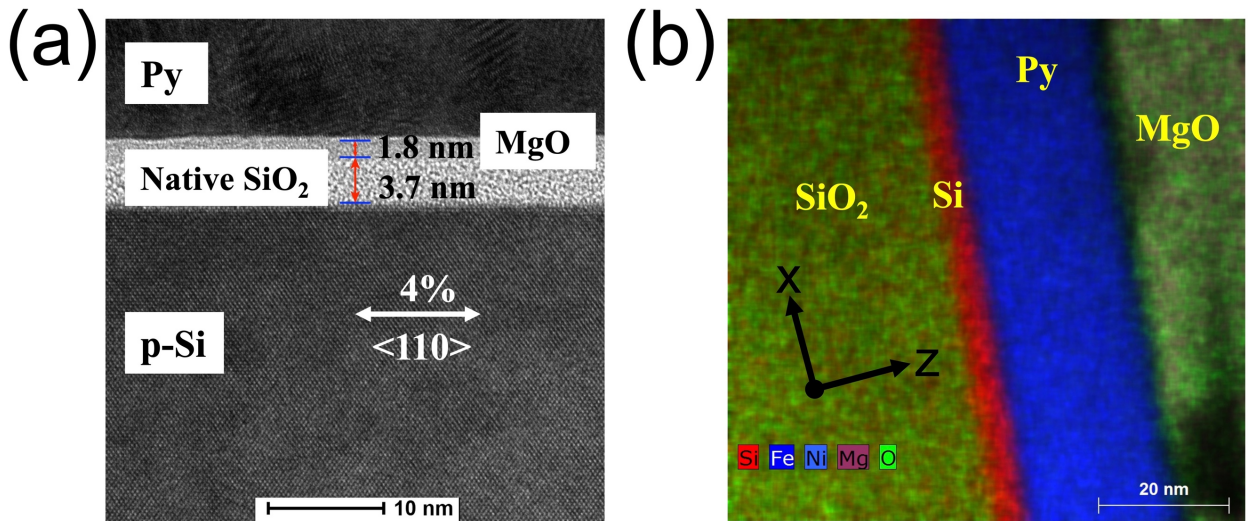

Figure 8. (a) Transmission electron micrograph showing the interfacial structure in a Py/MgO/native SiO$_2$/p-Si sample(Adapted and reproduced with permission from reference [28]. Copyright (2020) by the American Physical Society), and (b) Energy dispersive X-ray spectroscopy elemental map showing Py (25 nm)/p-Si (5 nm) sample (Adapted and reproduced with permission from reference [26]. Copyright (2025) with permission from Elsevier.). MgO in (b) is for electrical insulation of sample from heater element and not part of sample structure.

## 3 Interlayer charge transfer mediated behavior

Previous section presented the experimental evidence of flexoelectricity induced CE and interlayer charge transfer in two kinds of samples. The evidence was based on conductivity and Hall resistance behavior. However, the interlayer charge transfer will render the metal and Si layer to be electrically charged even though the heterostructure remains charge neutral; as stated earlier. This will not only change the electrical transport but it will also modify the electron-electron, electron-phonon and spin-phonon couplings in both the materials that form the heterostructure. This change is reflected in materials behavior and give rise to correlated electron behavior. In this section, the effects of interlayer charge transfer on the physical behavior of materials are discussed.

3.1 Electronic dynamical multiferroicity

The interlayer charge transfer gives rise to a gradient of charge carrier concentration with higher values near the interface and no change in charge carrier concentration far away from interface. For bulk materials, this gradient will be restricted to the interface or surface but this gradient will extend through the thickness in case of micro/nano scale thin films. This charge carrier concentration gradient induces a polarization with opposite direction in both the metal and degenerately doped Si layer. This polarization was called as flexoelectronic polarization as stated earlier since it was an electronic response to an externally applied strain gradient. In ferroelectric materials, the superposition of ferroelectric polarization and optical phonons (time-dependent rotational electric polarizations) generate temporally varying magnetic moments, which is called dynamical multiferroicity. Similarly, the superposition of flexoelectronic polarization and time dependent electric polarization of optical phonons give rise to electronic dynamical multiferroicity[35-39]. This behavior is described by following equation:

$$\boldsymbol{M}_t \sim \boldsymbol{P}_{F-El} \times \partial_t \boldsymbol{P}_{F-El} \qquad (1)$$

where $\boldsymbol{M}_t$, $\boldsymbol{P}_{F-El}$ and $\partial_t \boldsymbol{P}_{F-El}$ are respectively temporal magnetic moment, flexoelectronic polarization and time dependent polarization from optical phonons. This equation also describes the origin of new electron-phonon, spin-phonon and magnetoelectronic coupling, which will modify the underlying material behavior. The uncompensated interlayer charge transfer had been reported to induce ferroelectricity in layered 2D materials across different materials[40-43] except in present case, the origin of the multiferroicity is electronic.

3.1.1 Experimental evidence

3.1.1.1 Metal/degenerately doped Si -

The electronic dynamical multiferroicity was first reported in freestanding Pt (10 nm)/MgO/degenerately doped p-Si heterostructure sample[23]. However, the original work hypothesized that flexoelectronic polarization was due to charge separation instead of interlayer charge transfer, which was discovered later on. Electronic dynamical multiferroicity was discovered using angle dependent magnetoresistance (ADMR) for the magnetic field acting in the cross-sectional (110) plane of the p-Si layer in the sample as shown in Figure 9(a-c)[23]. The ADMR values were larger when the applied magnetic field was aligned along the temporal magnetic moment direction. The principal directions for Si (110) cross-sectional plane are shown in Figure 9(d). The flexoelectronic polarization was aligned along the two <112> directions (highest packing density) whereas the time dependent electric polarization of optical phonons was aligned along the <110> as shown in Figure 9(e-f). This gave rise to temporal magnetic moment acting along four <111> directions in the (110) cross-sectional plane of the p-Si layer as shown in Figure 9(e-f). Further, the temporal magnetic moment disappears at 100 K due to freezing of optical phonons, which validated the electronic dynamical multiferroicity. The polycrystalline Pt thin film cannot give rise to direction dependent response reported in this study.

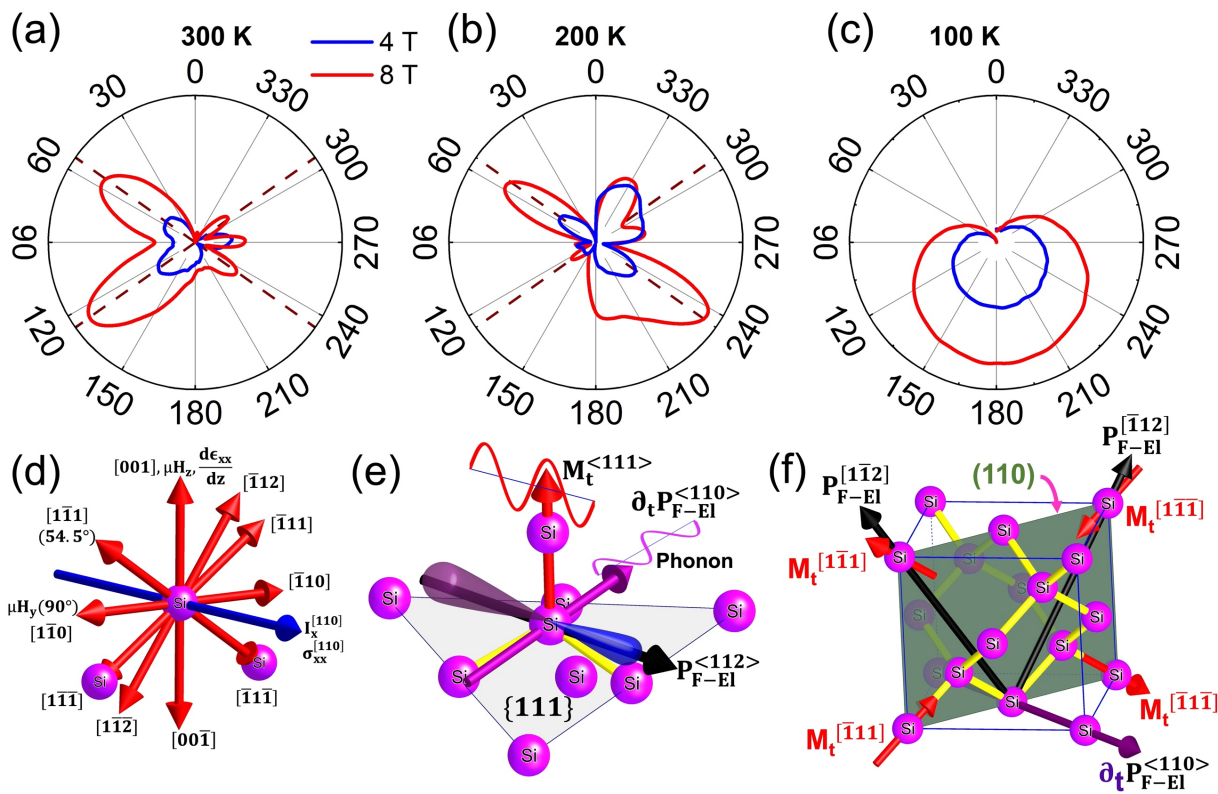

Figure 9. The polar plot of angle dependent (in the zy-plane) magnetoresistance (magnitude) in the Pt/p-silicon sample as a function of the constant magnetic field of 4 T and 8 T at (a) 300 K and (b) 200 K and (c) 100 K; the dotted lines represent $<111>$ directions. (d) The primary crystallographic directions (red color) in the zy- or (110) plane in the silicon sample observed in the measurement and direction of the current and stress (blue color), (e) a schematic showing the origin of electronic dynamical multiferroicity and the temporal variation in magnetic moment due to coupling between flexoelectronic polarization and dynamical polarization caused by optical phonons in case of Si, and (f) a schematic of the Si lattice showing the flexoelectronic polarization and the observed magnetic moments in the (110) cross-sectional plane caused by electronic dynamical multiferroicity. (Adapted and reproduced with permission from reference [23]. Copyright (2021) by the American Chemical Society).

3.1.1.2 Dielectric/degenerately doped Si-

The dielectric/degenerately doped Si experiment was done using MgO/n-Si bilayer structure as described previously and as shown in Figure 6(a-b)[23]. The interlayer charge transfer gave rise to anomalous Hall effect (AHE) response in the measurement at 5 mA as shown in Figure 6(b). The saturation magnetic field for AHE was 0.7 T, which corresponds to a magnetic moment of 1.2 $\mu_B$/atom[23]. This was an explicit example of large magnetic moment due to electronic dynamical multiferroicity. The saturation magnetization and AHE response was negligible at 1 mA of current due to smaller flexoelectronic effect and interlayer charge transfer.

3.1.2 Spin-phonon coupling

The electronic dynamical multiferroicity give rise to spin-phonon coupling as described by equation 1. The electronic dynamical multiferroicity mediated spin-phonon coupling was experimentally reported using thermal conductivity measurements using self-heating 3ω method[44] in a freestanding Py (25 nm)/MgO/degenerately doped p-Si (2 µm) sample[27, 45]. It was noted that Py layer did not contribute significantly to the thermal conductivity measurements because Si layer is three orders of magnitude more thermally conducting. The thermal transport in Si is phonon mediated. Due to interlayer charge transfer and electronic dynamical multiferroicity, the thermal conductivity of Si was found to be a function of magnetic field as shown in Figure 10(a-d). The interlayer charge transfer also led to interlayer magnetic coupling because thermal conductivity showed a valley corresponding to the saturation magnetization of Py layer. Originally, this behavior was attributed to spin dependent phonon scattering due to spin-phonon coupling. But, the mechanistic origin of spin-phonon coupling was not understood. The spin-phonon coupling arose from electronic dynamical multiferroicity. The thermal conductivity behavior changed as the temperature was reduced below 100 K because of phonon freezing at lower temperature, which confirms the origin of spin-phonon coupling from electronic dynamical multiferroicity. The spin dependent behavior did not occur in the absence of interlayer charge transfer. This spin-phonon coupling was also the underlying cause of AHE response described in previous section. The emergence of strong spin-phonon coupling in Si has significant consequences including spin-Hall effect, which will be discussed in the next subsection.

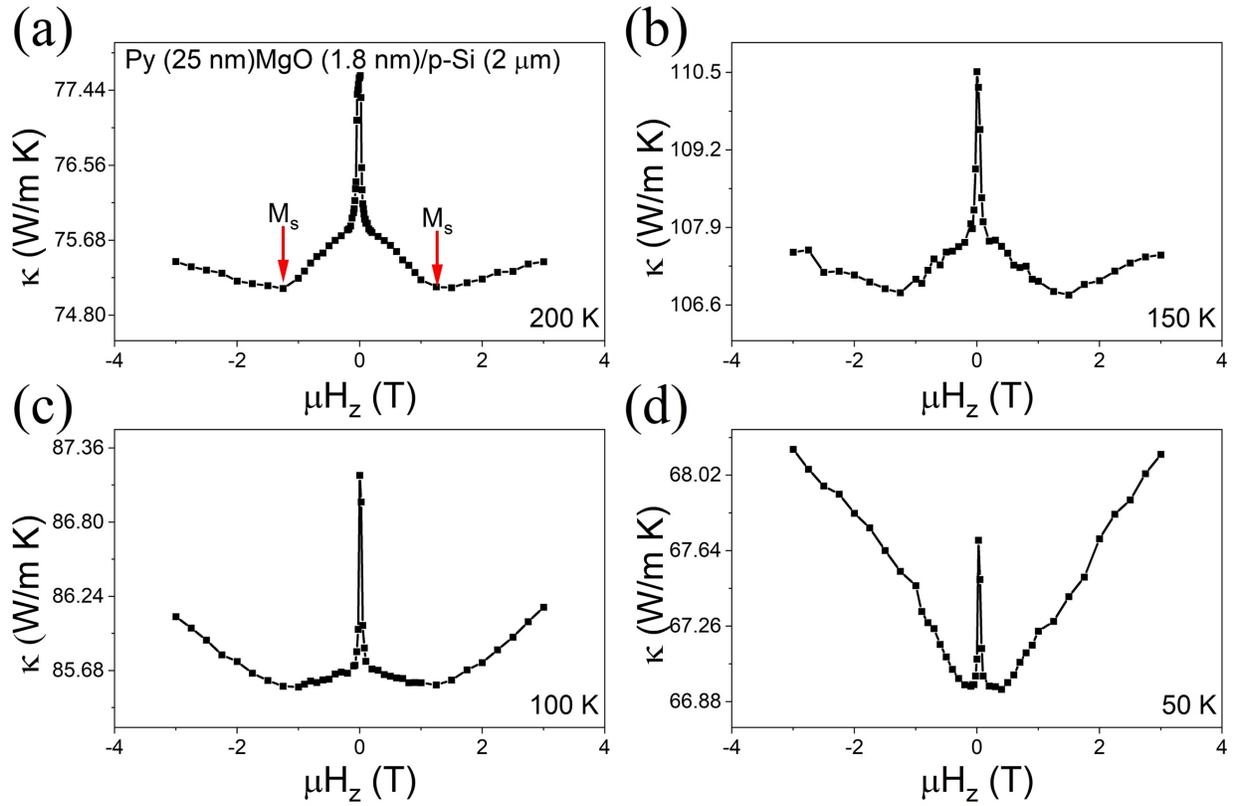

Figure 10. The thermal conductivity in Pd (1 nm)/ Py (25 nm)/MgO (1.8 nm)/p-Si (2 μm) as a function of magnetic field measured at (a) 200 K, (b) 150 K, (c) 100 K and (d) 50 K. (Reproduced with permission from reference [27]. Copyright (2018) with permission from Elsevier.).

3.1.3   Magnetoelectronic coupling

The flexoelectronic polarization is proportional to the gradient of the charge carrier concentration ($\mathbf{n'}$). Hence, the equation 1 can be rewritten as:

$$\mathbf{M}_t \propto \mathbf{n'} \times \partial_t \mathbf{n'} \qquad (2)$$

The equation 2 also describes cross-correlation between the magnetic properties (temporal magnetic moment) and the electronic properties (charge carrier concentration). If an external magnetic field (**B**) is applied, it will modulate the charge carrier concentration; as described by following equation:

$$(\mathbf{M}_t \pm \mathbf{B}) \propto (\mathbf{n} \pm \mathbf{\Delta n})' \times \partial_t (\mathbf{n} \pm \mathbf{\Delta n})' \qquad (3)$$

where **B** and **Δn** are external magnetic field and change in charge carrier concentration, respectively. This equation describes the magnetoelectronic coupling because an external magnetic field modulates the electronic properties. The magnetoelectronic coupling was demonstrated using Hall effect measurement. In the Hall effect measurement, the modulation in charge carrier concentration give rise to oscillatory Hall effect as shown in Figure 11(a-b). The magnetic period of the oscillation describes the temporal magnetic moment, which was reported to be 1.12 T in a freestanding MgO (1.8 nm)/degenerately doped p-Si (400 nm) sample as shown in Figure 11(b).

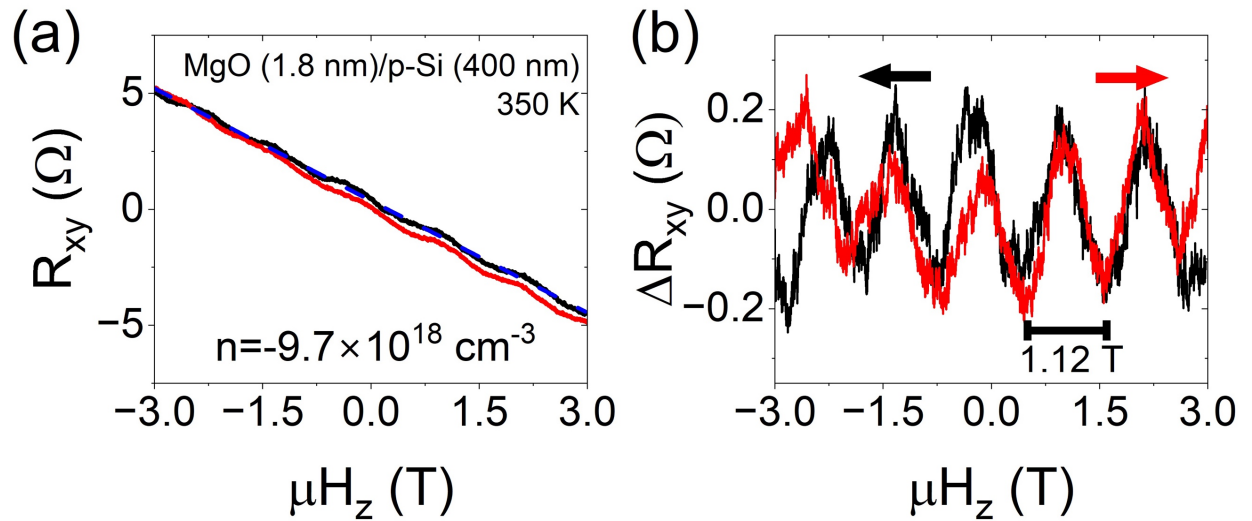

Figure 11. (a) The Hall response measured in MgO (1.8 nm)/p-Si (400 nm) sample at 350 K and (b) the oscillatory component of the Hall response showing a period of 1.12 T. (Adapted and reproduced with permission from reference [25]. Copyright (2024) by the American Physical Society).

The oscillatory Hall effect was first reported by Barklie and Pippard[46] in single crystal Sn at 4.2 K when a magnetic field acts along (or near) [001] axis. The oscillatory response was attributed to the migration of electrons due to magnetic breakdown in $\epsilon_1^5$ and $\delta_1^1$ orbits[46]. The oscillatory Hall resistance was described as:

$$\Delta n_a = 2.15 \times 10^{-4} \left(\frac{BI}{d}\right) \Delta V_H / \bar{V}_H^2 \qquad (4)$$

where $n_a$, B, I, d and $V_H$ are the excess electron per atom, magnetic field, sample current, diameter of the sample and Hall voltage, respectively. The amplitude of the oscillatory Hall response increased with increase in applied magnetic field. Whereas the magnitude of the observed response in p-Si was constant for increasing magnetic field. The frequency of the oscillations in case of Sn was periodic in 1/B whereas a constant period of ~1.12 T was reported in case of p-Si. Further, the oscillatory Hall response in p-Si sample was reported at 350 K as opposed to 4.2 K in case of Sn, as stated earlier. This showed that equation 3 completely described the oscillatory Hall response in p-Si sample and electronic dynamical multiferroicity was the underlying cause of magnetoelectronic coupling.

~~So far, the interlayer charge transfer mediated electronic dynamical multiferroicity behavior was demonstrated that did not exist prior to it. The evidence of modified spin-phonon, electron-phonon and magnetoelectronic coupling in the material were demonstrated, which arose from electronic dynamical multiferroicity.~~

## 3.2 Spin-Hall effect of charge carriers

The electronic dynamical multiferroicity from interlayer charge transfer give rise to spin phonon coupling. The spin-phonon coupling give rise to spin dependent electron-phonon scattering or phonon skew scattering of electrons. As a consequence, spin Hall effect response was observed across multiple experiments[28, 32, 33]. In the Py (25 nm)/MgO/degenerately doped p-Si (2 μm) freestanding sample[28] showed a spin-Hall magnetoresistance (SMR) response as a function of current as shown in Figure 12(a). The SMR response increased as the applied current increased similar to previously reported studies. The SMR response vanished as the temperature was lowered to 200 K as shown in Figure 12(b) and only anisotropic magnetoresistance (AMR) response from the Py layer was observed[23]. This behavior was consistently reported for p-Si as well as n-Si samples. These studies attributed the response to phonon skew scattering and spin-phonon coupling but the interlayer charge transfer mediated electronic dynamical multiferroicity behavior was not known at that time.

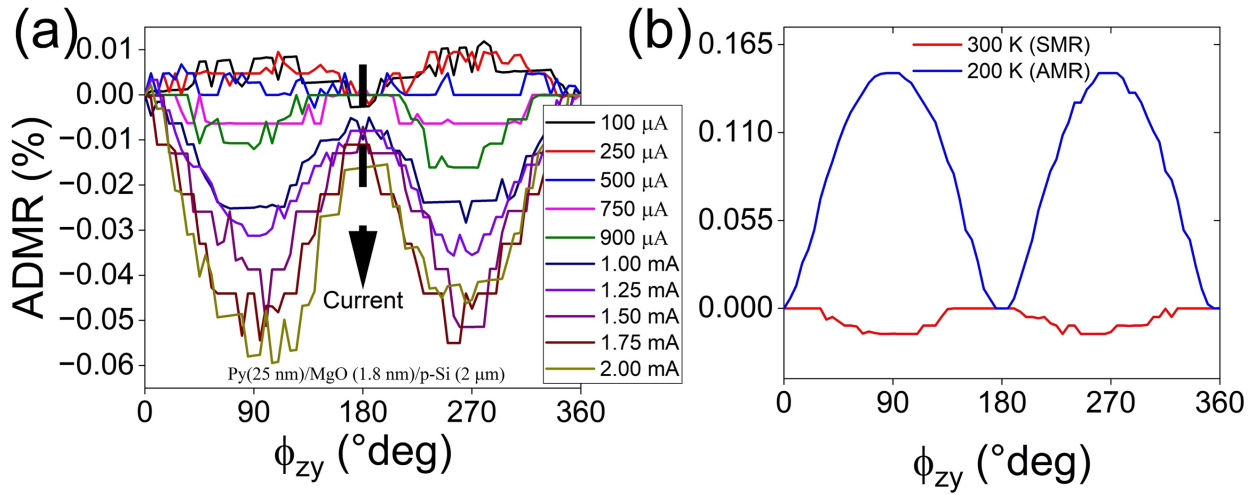

Figure 12. The ADMR response in Py (25 nm)/MgO (1.8 nm)/degenerately doped p-Si (2 μm) sample (a) as a function of applied current from 100 μA to 2 mA showing increase in SMR response

(Adapted and reproduced with permission from reference [28]. Copyright (2020) by the American Physical Society). And (b) the ADMR response as a function of temperature showing absence of SMR at 200 K (Adapted and reproduced with permission from reference [23]. Copyright (2021) by the American Chemical Society).

The SMR studies estimated the spin-Hall angle using conducting bimetallic configuration[47]. The spin-Hall angle values were orders of magnitude larger than reported for Si and similar to topological materials[28, 32, 33]. However, these reported values may not be applicable to the Si because interlayer charge transfer was not considered. The interlayer charge transfer between a magnetic material and non-magnetic material will have charge carrier current in the absence of magnetic field due to randomized domains as shown in Figure 13 (a). However, the interlayer charge transfer from a magnetic material will be spin-polarized when magnetic field is applied as shown in Figure 13 (b), which will induce a spin current across the interface. Hence, there will be two spin current in the Py/Si heterostructure: first transverse spin current is inside the Si layer due to spin-phonon coupling ($j_s^{s-ph}$) and second spin current is due to interlayer charge transfer ($j_s^{ICT}$) as shown in Figure 13(c). Total transverse spin current can be described by following equation:

$$j_s^{total} = j_s^{s-ph} \pm j_s^{ICT} \qquad (5)$$

The $j_s^{ICT}$ is a function of applied current because interlayer charge transfer increases with increase in strain gradient, which is the underlying cause of larger SMR response at larger current[28] as shown in Figure 12 (a). The $j_s^{s-ph}$ from spin dependent scattering is not a function of electric current through the sample. The spin to charge conversion is expected from spin-phonon coupling and phonon skew scattering.

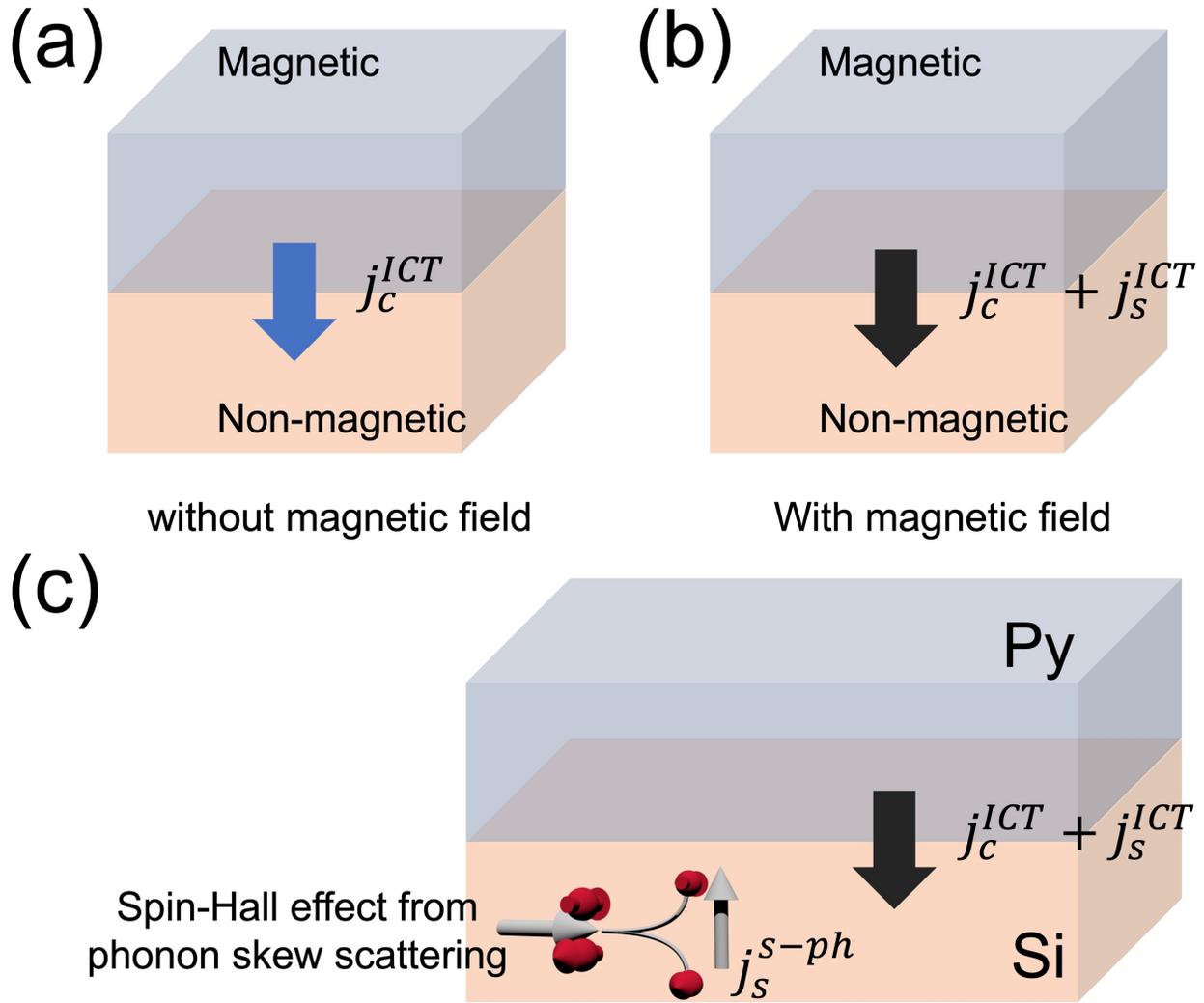

Figure 13. (a) Schematic showing the interlayer charge transfer current between magnetic and non-magnetic material in the absence of magnetic field, (b) schematic showing interlayer charge transfer accompanied by a spin current between magnetic and non-magnetic material under an applied magnetic field, and (c) schematic showing two spin currents from spin-Hall effect due to spin-phonon scattering and interlayer charge transfer in Py/Si heterostructure.

Current mathematical models for spin-charge interconversion do not consider interlayer charge transfer mediated spin current. Hence, the spin-Hall angle values reported in case of Py/Si heterostructures[28, 32, 33] are not correct and new mathematical models are needed to explain the SMR response. We speculate that interlayer charge transfer mediated spin current could be the primary reason for SMR like response but further research is required to establish that. In the spintronics, the spin current was not considered controllable using any external parameter since spin-Hall angle is a material property[48-51]. However, the spin current from interlayer charge transfer can be modified using strain gradient. This adds a new knob that can be controlled to enhance the required response for efficient spintronics (especially Si spintronics[52]) and spin-caloritronics[53] applications.

3.3  Magnetoelectronic electromagnon

The electronic dynamical multiferroicity arises due to superposition of flexoelectronic polarization from interlayer charge transfer and circularly polarized optical phonons. This superposition led to transformation of phonons to magneto-active phonons[24]. These magneto-active phonons are called as magnetoelectronic electromagnon that are described as:

$$\boldsymbol{M}_t^{\pm <hlk>} \propto \boldsymbol{P}_{F-El} \times \partial_t \boldsymbol{P}_{F-El}^{\pm} \propto \boldsymbol{n}' \times \partial_t(\boldsymbol{n}')^{\pm} \qquad (6)$$

Unlike magnetoelectric electromagnon observed in insulating multiferroics, the magnetoelectronic electromagnon arises in conducting material systems and couple to electronic charge as opposed to ionic charge.

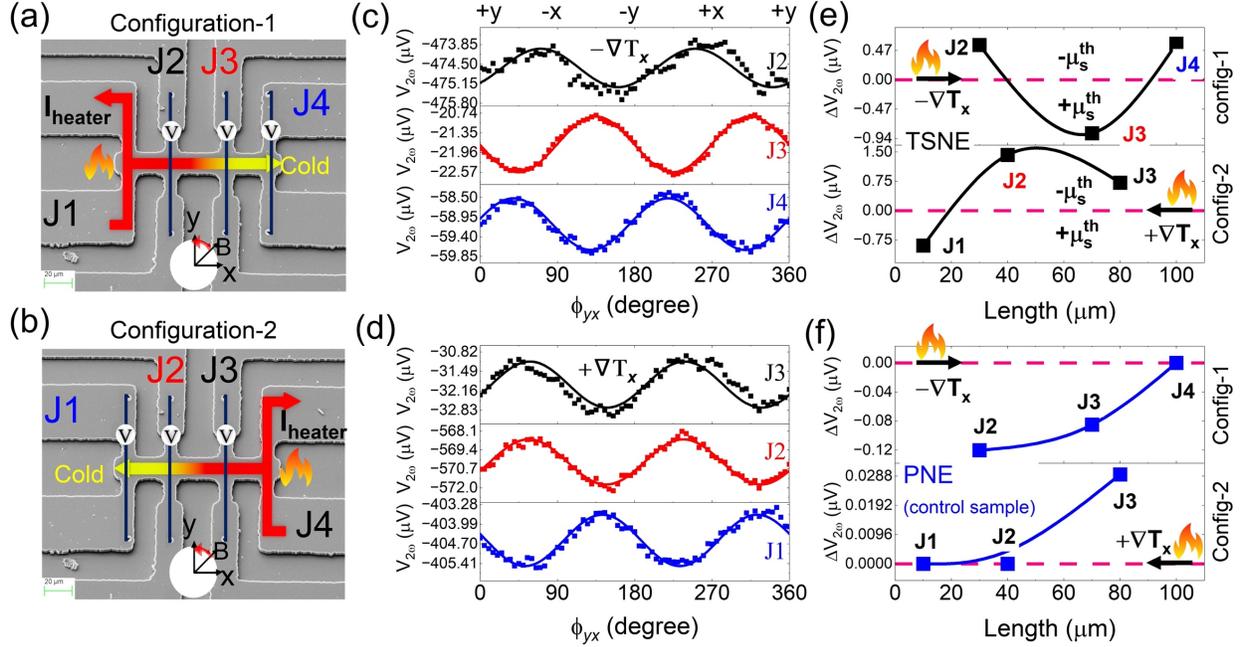

Figure 14. The scanning electron micrograph showing the representative device structure and experimental scheme for (a) configuration 1 and (b) configuration 2. The angle-dependent transverse magneto-thermoelectric response in Py (25 nm)/MgO (1.8 nm)/degenerately doped p-Si (2 μm) sample in the yx-plane at a constant applied magnetic field of 1 T (c) measured at J2, J3 and J4 in Configuration 1 and (d) measured at J3, J2 and J1 in Configuration 2. (e) The longitudinal distribution of the magnitude of transverse spin-Nernst effect for config 1 and config 2 with possible distribution of spin accumulation. (f) The planar Nernst effect response in control Py (25 nm) sample showing the longitudinal distribution and response diminishes farther from the heat source. The solid line in (e) and (f) for representation only. (Reproduced with permission from reference [24]. Copyright (2023) by the American Physical Society).

The magnetoelectronic electromagnon carry heat as well as spin. As a consequence, the thermal transport will lead to spin accumulation along the direction of propagation in addition to transverse spin current from spin dependent phonon scattering. This behavior was experimentally reported in a freestanding Py (25 nm)/MgO/degenerately doped p-Si sample shown in Figure 14 (a)-(b). The sample had four Hall bars. The magneto-thermal measurement was taken in two configurations: (1) left most Hall junction was heated by passing current and angle dependent transverse second harmonic response was measured in other three Hall junctions and (2) the heating was done at right most Hall junction and response was measured in the other three Hall junctions. A spatially varying transverse spin-Nernst effect (TSNE) response was reported in both configuration where sign of the TSNE response changed along the length as shown in Figure 14 (c-e). In the (110) cross-sectional plane of p-Si, the magnetoelectronic electromagnon with temporal magnetic moment $\mathbf{M}_t^{[1\bar{1}\bar{1}]}$, $\mathbf{M}_t^{[1\bar{1}1]}$, $\mathbf{M}_t^{[\bar{1}11]}$ and $\mathbf{M}_t^{[\bar{1}1\bar{1}]}$ were reported. The thermal spin accumulation ($\mu_s^{th}$) at any given spatial location can be described as:

$$\mu_s^{th}(x,y) \propto \mathbf{M}(x,y) \propto \sum \mathbf{M}_t^{<hlk>} = \mathbf{M}_t^{[1\bar{1}\bar{1}]} + \mathbf{M}_t^{[1\bar{1}1]} + \mathbf{M}_t^{[\bar{1}11]} + \mathbf{M}_t^{[\bar{1}1\bar{1}]} \qquad (7)$$

Hence, the spin accumulation with spin-up or spin-down configuration will be a function of spatial coordinates, which gave rise to the observed spatially varying TSNE responses. This behavior was in contrast to the conventional Planar Nernst effect (PNE) measurement in a control sample as shown in Figure 14 (f). This experiment demonstrated long distance (~100 µm) spin transport. Similar behavior was reported in case of Pt (10 nm)/MgO (1.8 nm)/degenerately doped p-Si (2 µm)[24], MgO (1.8 nm)/degenerately doped p-Si[54] and MgO (1.8 nm)/degenerately doped n-Si samples[54] as well. The mechanistic origin of the long-distance spin transport in MgO (1.8 nm)/degenerately doped p-Si[54] and MgO (1.8 nm)/degenerately doped n-Si samples[54] was absent in the original published article, which can be understood in framework of flexoelectricity mediated CE and interlayer charge transfer.

These studies showed the longest (>100 µm) spin angular momentum transport without any significant dissipation in both metal/Si and dielectric/Si heterostructures. It was an order of magnitude longer than reported values in the case of quantum spin-Hall state[55], quantum-Hall antiferromagnet[56], antiferromagnetic insulators[57-59], magnetic insulators[60] and spin-superfluidity[61]. The long-distance spin transport coupled with large spin-Hall effect can make the spintronics devices a reality. As stated previously, the magnitude of the spin current can be increased using strain gradient mediated interlayer charge transfer. All these studies must be further explored for the application in spin-based quantum systems.

3.4     Topological Hall effect of magnetoelectronic electromagnon

A freestanding thin films structure under an applied strain gradient is equivalent to a gradient index medium since a strain field will exist perpendicular to the magnetoelectronic electromagnon transport direction (longitudinal direction of the sample). Magnetoelectronic electromagnon are analogues to an electromagnetic ray (photon) as shown in Figure 15. The deflection of a ray in an inhomogeneous medium is given by:

$$\delta \mathbf{r}_{tc} = -\sigma_c \lambda_{t0} \int_C \frac{\mathbf{p}_t \times d\mathbf{p}_t}{p_t^3} = -\sigma_c \lambda_{t0} \frac{\partial \Theta^B}{\partial \mathbf{p}_{tc}^{(0)}} \qquad (8)$$

where $\sigma_c$, $\lambda_{t0}$, $\mathbf{p}_t$ and $\Theta^B$ are helicity, wavelength, momentum and Berry phase[62], respectively.

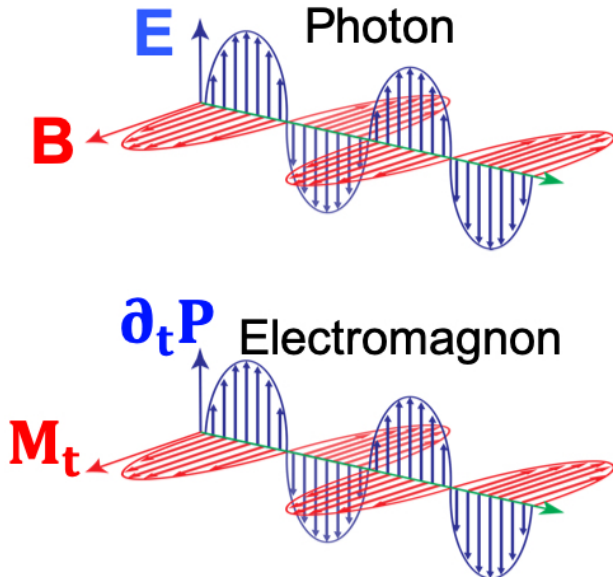

Figure 15. Schematic showing that the electromagnon are analogues to photon. (Adapted and reproduced with permission from reference [24]. Copyright (2023) by the American Physical Society)

This behavior was experimentally tested and the topological Berry phase was reported in case of magnetoelectronic electromagnon transport in inhomogeneously strained MgO/p-Si (2 μm) thin film. The topological behavior was uncovered using Hall resistance and second harmonic Hall voltage as shown in Figure 16(a-b). The Hall resistance did not exhibit the change in charge carrier concentration as the measurement current was increased as shown in Figure 16(a), unlike other studies. The second harmonic Hall voltage arose from Nernst effect, which appeared as a linear response as a function of magnetic field. However, the Nernst response showed a characteristic feature of topological Berry phase as shown in Figure 16(b) having peak at around 0.6 T[24]. It was called as topological Nernst effect (TNE) response[24]. This measurement clearly demonstrated the topological Hall effect of magnetoelectronic electromagnon due to inhomogeneous strain. Further, the magnetic moment of magnetoelectronic electromagnon also proved the electronic dynamical multiferroicity since Nernst effect arises from electron-phonon coupling.

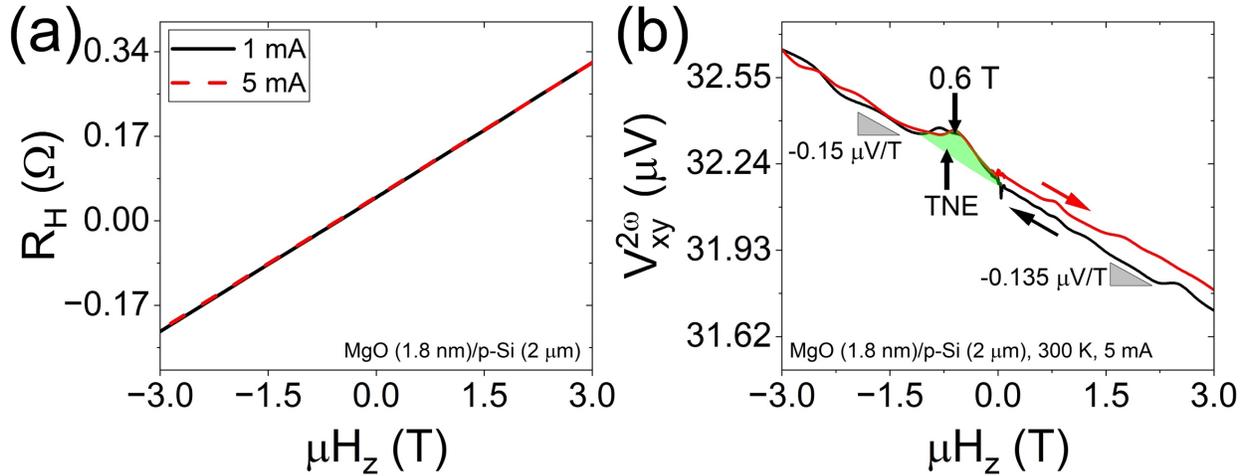

Figure 16. (a) The Hall response measured at 300 K in MgO (1.8 nm)/degenerately doped p-Si (2 mm) for 1 mA and 5 mA of applied current bias and a magnetic field sweep between 3 T and -3 T and (b) the second harmonic Hall response at 300 K for an applied current bias of 5 mA showing possible TNE behavior. The response in the green shaded region is expected to arise from topological Nernst effect. (Adapted and reproduced with permission from reference [24]. Copyright (2023) by the American Physical Society).

In another experiment, the magnetoelectronic electromagnon having momentum along the [110] and [$\bar{1}\bar{1}$0] directions deflected to the opposite longitudinal edges of the freestanding Py (25 nm)/MgO (1.8 nm)/degenerately doped p-Si sample[24]. It was direct-evidence of spin-Hall effect of magnetoelectronic electromagnons. In addition, the magnetoresistance measurement shows accumulation of spin-up and spin-down charge carriers on opposite sample edges, which give rise to opposite bumps due to spin-canting in the measurement as shown in Figure 17 (a-b). The edge dependent longitudinal resistances were measured to be $R_{left}$=19.3 Ω and $R_{right}$=18.59 Ω at 300 K at zero field, which showed transverse asymmetry. This response was due to scattering from magnetoelectronic electromagnon as shown in Figure 17 (a-b)[24], which demonstrated the combined effect from spin-Hall effect of charge carriers and topological Hall effect of magnetoelectronic electromagnons.

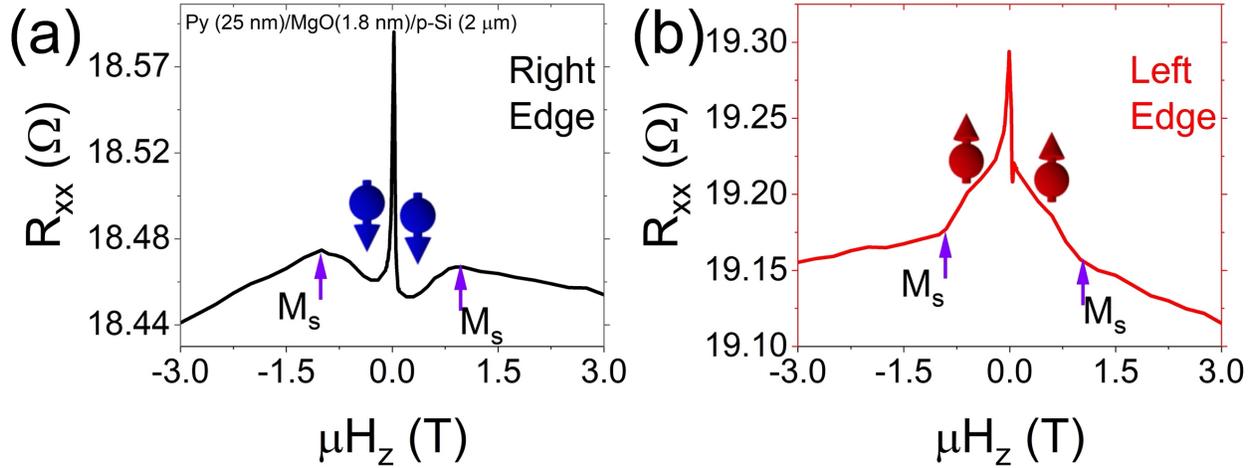

Figure 17. The edge dependent magnetoresistance measured for an applied magnetic field sweep from 3 T to -3 T (a) right edge and (b) left edge in the Py (25 nm)/MgO (1.8 nm)/ degenerately doped p-Si sample at 300 K. The $M_s$ shows the saturation magnetization corresponding to Py (Adapted and reproduced with permission from reference [24]. Copyright (2023) by the American Physical Society).

3.5   Inhomogeneous magnetoelectronic multiferroic effect

The magnetoelectronic electromagnon gave rise to spatially modulated spin accumulation in thermal transport measurements as shown previously. The spatially modulated spin behavior was also expected to arise in charge transport measurements. The spatially modulated spin behavior study was reported in a sample structure consisted of Py (25 nm)/MgO (1.8 nm)/ degenerately doped p-Si (400 nm). The device structure was similar to the one previously shown in Figure 14 (a-b). The Hall resistance measurements were taken at two Hall junctions (J2 and J3) as shown in Figure 18 (a). The charge carrier concentration for positive and negative magnetic field was interpreted as spin-up and spin-down concentrations, respectively, as shown in Figure 18 (a). The charge carrier concentration showed a transition where spin-down concentration was higher at Hall junction J2 as compared to spin-up and vice versa at Hall junction J3 40 µm away as shown in Figure 18 (b). The average charge carrier density was larger in measurement at Hall junction J2 as compared to the one at Hall junction J3. Further, the anomalous Hall resistance was larger at Hall junction J3 than at Hall junction J2 as shown in Figure 18 (a). These spatially modulated charge carrier, spin concentrations and anomalous Hall resistance further supported the conclusions from equation 5 and previous thermal transport measurements. This behavior was called inhomogeneous magnetoelectronic multiferroic effect; analogues to the inhomogeneous magnetoelectric effect since spatial magnetic inhomogeneity in a magnetic crystal[63-67] give rise to electric polarization and vice versa[39]. Similar spatial coordinate dependent behavior was reported in Py (25 nm)/MgO (1.8 nm)/ degenerately doped p-Si (2 µm) sample as well[24].

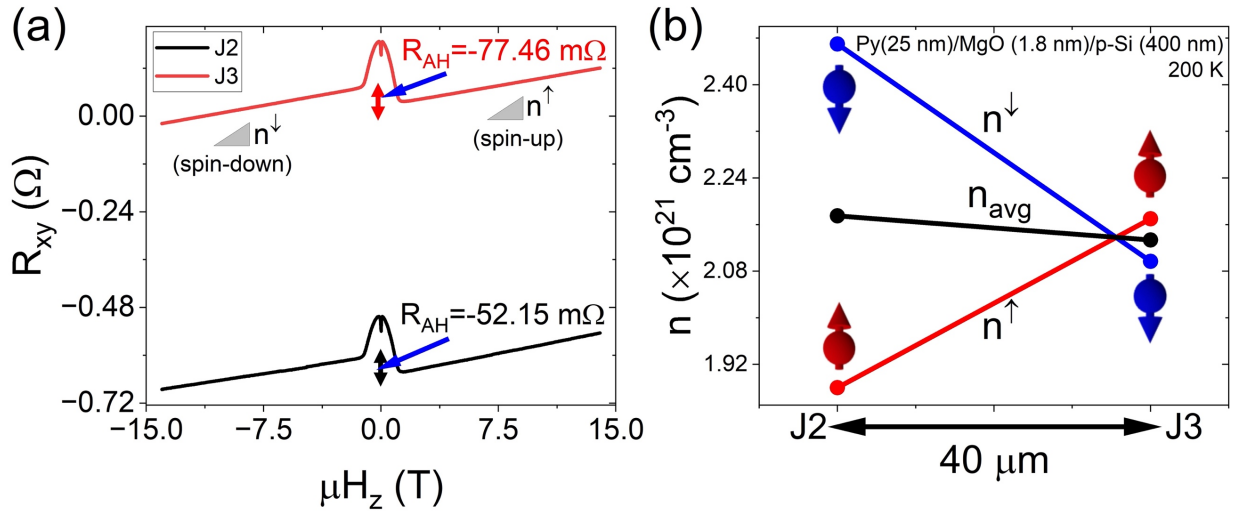

Figure 18. (a) the Hall effect responses at Hall junctions J2 and J3 in a Py (25 nm)/MgO (1.8 nm)/p-Si (400 nm) sample measured at 200 K for an applied magnetic field from 14 T to -14 T, and (b) the longitudinal modulations in the charge carrier concentration and spin density.

(Reproduced with permission from reference [24]. Copyright (2023) by the American Physical Society).

The longitudinal and transverse modulation of the spin and charge in real space can give rise to multiple physical behavior. The incommensurate spin-density wave from the superposition of magnetoelectronic electromagnons[68, 69] of opposite chirality is one of them. Similarly, the incommensurate charge density wave may also arise since charge carrier density is also a function of longitudinal and transverse spatial coordinates. The interlayer charge transfer can potentially lay the foundation of engineering correlated electron systems using conducting heterostructures, which requires a significant cross-domain research effort.

3.6    Non-reciprocity of magnetoelectronic electromagnon

The non-reciprocity of circularly polarized phonons is an essential condition for the previously described spatially varying spin accumulation The broken inversion symmetry due to strain gradient in the samples will lift the degeneracy of optical phonons and, as a consequence, magnetoelectronic electromagnon, which gave rise to non-reciprocity $(\partial_t(\boldsymbol{n}')^+ \neq \partial_t(\boldsymbol{n}')^-)$. The non-reciprocity was verified using longitudinal second harmonic response as shown in Figure 19(a-b). Assuming behavior similar to magnetochiral anisotropy, the corresponding coefficient was estimated to be 0.151 $A^{-1}T^{-1}$ and 0.352 $A^{-1}T^{-1}$ at 300 K and 200 K, respectively[24]. The non-reciprocal response diminished at 100 K. The reported value of magnetochiral anisotropy in Si FET interfaces was reported to be 0.1 $A^{-1}T^{-1}$[70-72] at 2.92 V of gate bias. This indicated the large interlayer charge transfer and flexoelectronic polarization but quantification of it was not done. Further theoretical and experimental studies are required to address it.

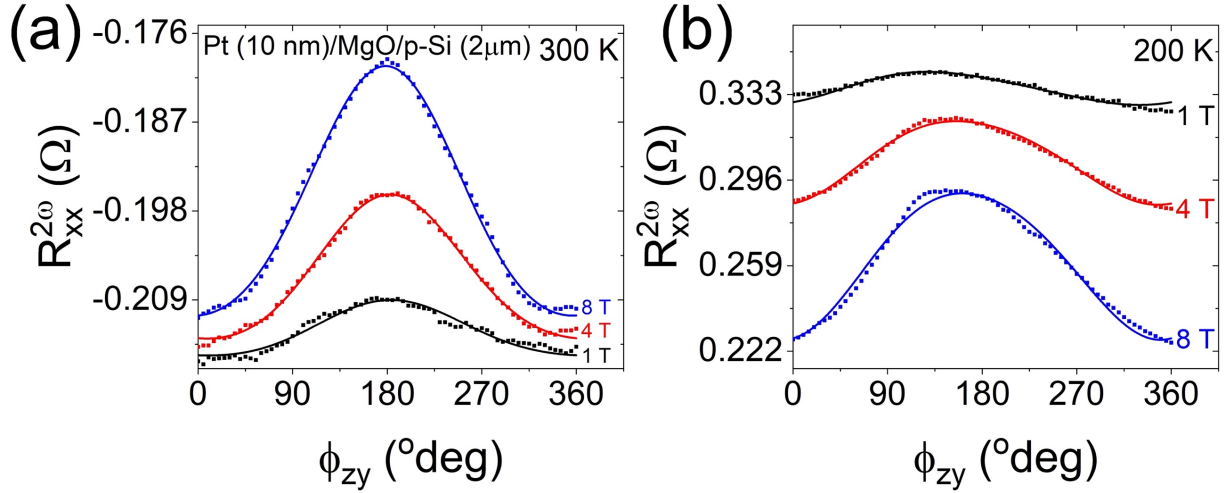

Figure 19. The angle dependent longitudinal second harmonic behavior for applied magnetic field of 1 T, 4 T and 8 T in Pt (10 nm)/MgO (1.8 nm)/degenerately doped p-Si (2 μm) sample at (a) 300 K and (b) 200 K. Solid line represent curve fit. (Adapted and reproduced with permission from reference [24]. Copyright (2023) by the American Physical Society).

3.7  Flexoelectronic proximity effect

The interlayer charge carrier transfer will give rise to a new proximity effect called as flexoelectronic proximity effect. The flexoelectronic proximity effect led to continuity of wave function across the interface, which induced spin-orbit coupling, exchange coupling and Berry curvature from one material to another. The flexoelectronic proximity effect was discovered in thermal measurements in Py (25 nm)/p-Si (5, 25, 50 and 100 nm) samples[26, 30, 31]. These measurements demonstrated exchange bias due to antiferromagnetic RKKY interlayer exchange coupling between Py and p-Si layer as shown in Figure 20 (a) [26]. The magnetic moment arose in p-Si due to interlayer charge transfer and flexoelectronic proximity effect[26]. It did not arise from electronic dynamical multiferroicity because the response persisted at low temperatures[26]. The antiferromagnetic RKKY interlayer coupling was reported in {Py(10 nm)/p-Si (25 nm)}$_3$ and

{Py(10 nm)/p-Si (10 nm)}$_3$ multilayer heterostructures as well[26]. All these samples were encapsulated using MgO layer on top and not freestanding, which showed the second method to induce strain gradient in the samples. Further, the control experiments showed that samples without strain gradient did not exhibit interlayer charge transfer and RKKY interlayer coupling[26].

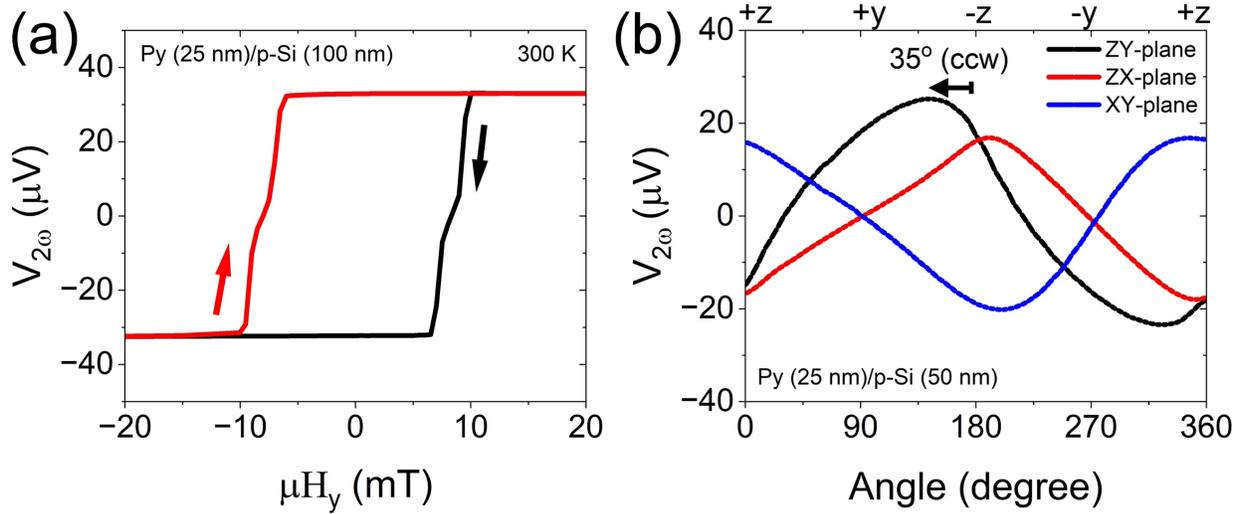

Figure 20. (a) magneto-thermoelectric transport measurement showing hysteresis with antiferromagnetic RKKY interlayer exchange coupling in a Py (25 nm)/p-Si (100 nm) sample (Adapted and reproduced with permission from reference [26]. Copyright (2025) with permission from Elsevier.), and (b) the angle dependent magneto-thermoelectric transport behavior in Py (25 nm)/p-Si (50 nm) sample showing 35° counter clockwise canting of magnetic moment(Adapted and reproduced with permission from reference [30]. Copyright (2025) with permission from AIP publishing.).

The interlayer exchange coupling in conjunction with induced spin-orbit coupling from Py layer gave rise to the proposed topological spin texture. The topological spin texture gave rise to angle dependent thermoelectric responses, as shown in Figure 20 (b), that were consistent with the behavior expected from hexagonally warped helical spin texture[73]. The canting of magnetic moment in two samples was found to be 35° counter clockwise as shown in Figure 20 (b) [30] and 45º clockwise [26]. Originally these responses were attributed to conventional spin-Seebeck effect[30, 31]. However, the discovery of interlayer charge transfer in Py/p-Si structures led to reanalysis of the mechanistic origin of the observed response to topological spin texture due to interlayer charge transfer mediated interlayer RKKY coupling and magnetic canting. The observed behavior is found in samples having strain gradient only. The observation of the topological spin texture suggests magnetic frustrated system[74, 75] in the Py/p-Si heterostructure samples, which could give a new direction to research in this field.

## 4    Future research directions and challenges

The research studies presented in this review demonstrated the flexoelectricity induced CE and interlayer charge transfer in conducting micro/nano scale thin film heterostructures. The reported behavior may lead to applications in spintronics, spin-caloritronics, straintronics, multiferroics, thermoelectrics and quantum devices. However, there is need to study interlayer charge transfer across materials and dimensional scale using both theoretical and experimental methods in order to realize this potential. Some of the challenges are already stated in the previous section. Here, we list a few more future directions that can help this research area grow.

4.1    Critical length scale

As shown in Figure 1 (c), a critical length scale must exist based on the material properties where interlayer charge transfer will not affect the bulk properties and charge accumulation will be considered interfacial only. In one of the reported studies, the Mott MIT was observed in the 400 nm thick p-Si layer, which suggests that the critical length scale is larger than 400 nm in case of p-doped Si. The electron mean free path in Si is order of magnitude smaller than 400 nm at $10^{19}$ cm$^{-3}$ charge carrier concentration[76]. Hence, the electron mean free path cannot be the critical length scale for interlayer charge transfer. Similarly, the length scale in the metal layer is also unknown. A combined theoretical and experimental approach is required to address this question.

4.2     Material selection

The studies presented in this review article utilized metal and degenerately doped Si. It was a good selection because their electronic properties were significantly different, which allowed a strong interlayer charge transfer response. There was a single study where metal-metal contact behavior was reported in freestanding Py (10 nm)/Au (100 nm) sample[77]. However, there was no explicit evidence of interlayer charge transfer even though thermal properties showed interlayer coupling. Future research must include metal-metal, metal-bad metal and other combination of conducting materials for interlayer charge transfer studies, which can, potentially, discover new condensed matter behavior. Further, the interfacial quality and materials along with their contribution to interlayer charge transfer is also needed to be explored.

4.3     Theoretical models for interlayer charge transfer

The studies presented in this review are experimental only and no theoretical study have been attempted to explore the interlayer charge transfer. The interlayer charge transfer in conducting micro/nano structures present a unique challenge. The interlayer charge transfer across the interface leaves one material positively and other negatively charged even though the heterostructure is charge neutral. Further, the gradient in the charge carrier distribution give rise to flexoelectronic polarization, as stated earlier. This requires new modelling methods to account for lack of charge neutrality. Further, the flexoelectronic proximity effect will lead to continuity of electronic wavefunction continuity across the interface, which is an essential ingredient for new physical behavior. All these challenges require new modelling methods.

The research demonstrated that interlayer charge transfer changed the electron-electron, electron-phonon, spin-phonon and magnetoelectronic coupling. However, theoretical models are needed to quantify the changes, which can also provide new methods to experimentally control them for applications.

4.4     Experimental setup

All the experimental studies reported and presented in this review are based on transport measurements. However, the sample structure used in the experimental studies could only be used to estimate the interlayer charge transfer. A direct measurement of interlayer charge transfer using transport measurements will require a new sample structure and experimental setup. In the layered heterostructure, only layer on the top can be accessed for spectroscopic techniques. Further, an encapsulated sample is completely hidden and spectroscopy methods cannot be utilized to study the materials behavior. These problems render current experimental setup unsuitable for experiments using other techniques (spectroscopy) and modification will be required to address this challenge.

4.5     Spectroscopy techniques-

As stated earlier, the spectroscopy techniques can provide additional information for both transport measurement and theoretical models. There was only one attempt to study the behavior using Raman spectroscopy[27]. The study was insufficient due to lack of understanding of interlayer charge transfer. However, the magneto-Raman spectroscopy can reveal extensive information regarding the electronic dynamical multiferroicity and temporal magnetic moment of optical phonons. The Raman spectroscopy can also shed light on renormalization of the phonons due to flexoelectronic polarization[78]. Atomic force microscope (AFM) has been used to study contact electrification[79] and can be utilized to study interlayer charge transfer. Magnetic force microscopy and KPFM can be used to identify magnetic and flexoelectronic polarization behavior, respectively. X-ray spectroscopy and THz spectroscopy[80] can be used to uncover the magnetoelectronic electromagnon spectra and behavior. The strain gradient coupled with interlayer charge transfer breaks the structural inversion asymmetry. As a consequence, the optical second harmonic generation can be used to interlayer charge transfer[81]. These are some of the spectroscopy techniques that can be applied to interlayer charge transfer but future research may not be restricted to these techniques only. These spectroscopy techniques can elaborate and quantify the electron-electron, electron-phonon, spin-phonon and magnetoelectronic couplings from interlayer charge transfer. As a consequence, theoretical models can be improved to reflect the observed behavior.

In conclusion, we have presented a summary of the current status of research in interlayer charge transfer from flexoelectricity induced CE in micro/nano scale heterostructures of conducting materials. The research in this field is at a nascent stage. It, currently, presents endless opportunities and challenges some of which we have listed here. Some of the physical behavior already discovered may need further studies to uncover their atomistic origin. There is a strong possibility of practical realization of electronic multiferroic materials using interlayer charge transfer. The electronic dynamical multiferroicity can, potentially, lead to quantum devices using magneto-active phonons or magnetoelectronic electromagnons at room temperature. An extensive cross domain research effort is need to solve the issues presented in this article in order to realize the full potential of CE induced interlayer charge transfer.